\documentclass[12pt,preprint]{aastex}
\shorttitle{Eta Carinae Event}
\shortauthors{Soker \& Hamann}

\def \s{~\rm{s}}
\def \km{~\rm{km}}

\def \K{~\rm{K}}

\def \AU{~\rm{AU}}
\def \erg{~\rm{erg}}

\def \yr{~\rm{yr}}

\begin{document}

\title{COMPARING ETA CARINAE WITH THE RED RECTANGLE}

\author{Noam Soker\altaffilmark{1}}

\altaffiltext{1}{Department of Physics, Technion$-$Israel
Institute of Technology, Haifa 32000 Israel;
soker@physics.technion.ac.il.}

\begin{abstract}
I compare the structures of the bipolar nebulae around the massive binary system
$\eta$ Carinae and around the low mass binary system HD~44179.
While $\eta$ Carinae is on its way to become a supernova, the Red Rectangle is
on its way to form a planetary nebula.
Despite the two orders of magnitude difference in mass, these two systems show
several similarities, both in the properties of the stellar binary systems and
the nebulae. From this comparison and further analysis of the accretion process
during the 20~years Great Eruption of $\eta$ Carinae, I strengthen the binary
model for the formation of its bipolar nebula$-$the Homunculus.
In the binary model a large fraction of the mass lost by the primary star
during the Great Eruption was transferred to the secondary star (the companion);
An accretion disk was formed around the companion, and the companion
launched two opposite jets.
I show that the gravitational energy of the mass accreted onto the secondary star
during the Great Eruption can account for the extra energy of the Great Eruption,
both the radiated energy and the kinetic energy in the Homunculus.
I also conclude that neither the proximity of the primary star in $\eta$ Car to
the Eddington luminosity, nor the rotation of the primary star are
related directly to the shaping of the Homunculus.
I speculate that the Great Eruption of $\eta$ Carinae was triggered by disturbance
in the outer boundary of the convective region, most likely by magnetic activity,
that expelled the outer radiative zone.

\end{abstract}
\section{INTRODUCTION}
\label{intro}

Many diverse astrophysical objects posses bipolar structures composed of
low density bubble pair, a narrow waist between the bubbles,
and where each bubble is fully or partially bounded by a thin dense shell.
Examples include clusters of galaxies, e.g. Perseus (Fabian et al.\ 2002),
and A 2052 (Blanton et al. 2001), symbiotic nebulae, e.g.,
He 2-104 (Corradi \& Schwarz 1995; Corradi et al. 2001),
planetary nebulae (PN), e.g.,  NGC 3587 (PN G148.4+57.0;
e.g., Guerrero et al.\ 2003), and the bipolar nebula of $\eta$ Car$-$the
Homunculus (e.g., Morse et al. 1998; {{{ Davidson et al. 2001; Smith 2006). }}}
The similarity between bipolar PNs and bipolar symbiotic nebulae was discussed
by Corradi \& Schwarz (1995), the similarity between bipolar PNs and bubble
pairs in clusters of galaxies was discussed in a series of papers
(Soker \& Bisker 2006 and references herein), and the similarity of
$\eta$ Car and its Homunculus to the other classes was discussed by
Soker (2001b, 2004, 2005b), Smith (2003) and {{{ Smith et al. (2005). }}}
The nuclear spectrum of the PN M2-9 was compared to that
of $\eta$ Car by Balick (1989) and Allen \& Swings (1972).

The similarity of the bipolar structures in the different classes
hints on a common shaping mechanism (Soker 2004).
In clusters of galaxies such bubbles are known to be formed by
opposite jets, ejected by an accretion disk, which are detected by radio emission
(e.g., Hydra A; McNamara et a.\ 2000); the jets expand into a previously
existing medium.
The only source of angular momentum sufficient to form accretion disks, around the
primary or (more likely) around the secondary star,  in evolved
stars is the orbital angular momentum of a stellar (or in some cases substellar)
companion (Soker 2004; Soker \& Livio 1994; Morris 1987; Soker \& Bisker 2006).
This led me to suggest that during the Great Eruption the secondary star in
$\eta$ Car accreted mass, formed an accretion disk, and for the
$\sim 20 \yr$ (1837-1856) duration of the Great Eruption blew two jets that shaped
the Homunculus (Soker 2001b; 2004).

Nota et al. (1995) noted that many nebulae around luminous blue variables (LBVs)
posses axisymmetrical structure (see also O'Hara et al. 2003).
Nota et al. (1995) also attributed the bipolar shape of the Homunculus to
the binarity of $\eta$ Car, but using a different model than I proposed.
They use the interacting wind model, where a very dense and slow equatorial
flow constraints the fast wind to form bipolar structure.
Not only this mechanism cannot work in clusters of galaxies and was found
unapplicable to PNs (Balick 2000; Soker \& Rappaport  2000),
it also cannot work in the case of $\eta$ Car (Dwarkadas \& Balick  1998; Smith 2006).

On the other side of the debate on the shaping mechanism of $\eta$ Car,
there are models for the shaping of the Homunculus by the
primary star itself (at most, the companion spins-up the primary, but
does not interact with the wind itself and does not blow a significant wind;
e.g., Langer et al. 1999; Maeder \& Desjacques 2001;
Dwarkadas \& Owocki 2002; Smith et al.\ 2003a; Gonzalez et al.\ 2004;
van Boekel et al.\ 2003; Matt \& Balick 2004).
In a previous paper (Soker 2004; also Soker 2005b) I listed several great
difficulties with single star models for the shaping of $\eta$ Car.
Ignoring these arguments, Smith (2006) used his nice observational
results to conclude that the secondary star could not have directly shaped
the Homunculus during the Great Eruption, and that the bipolar shape must
come from the rapid rotation of the primary star.
In this paper I dispute Smith's conclusions, and show that the mass distribution
he finds in $\eta$ Car supports the binary shaping model.

\section{ENERGY AND ANGULAR MOMENTUM CONSIDERATIONS}
\label{energy}

\subsection{Energy Budget}
The source of the kinetic energy of the expanding bipolar nebula in
the binary model is mainly the accretion energy onto the companion, and not the
orbital gravitational energy (Soker 2001b, 2004).
For that, I find the claim that the mass loss during the Great Eruption could not have been
redirected toward the poles by deflection from its companion star, because the amount
of kinetic energy in the polar ejecta is greater than the binding energy of the current
putative binary system (Smith 2006), {{{ unjustified. }}}

I will upscale quantities in the expression derived in Soker (2001b)
according to the new estimate of the Homunculus mass
(Smith et al. 2003; Smith 2006).
I take a main sequence binary companion of pre-outburst mass
$M_{20} \simeq 30 M_\odot$ and an initial radius of $R_{20} \sim 10 R_\odot$.
The primary star lost $M_{\rm GE-1} \sim 20 M_\odot$ in a more or less
spherical geometry.
Say $\sim 12 M_\odot$ were captured by the secondary, out of which
$M_{\rm GE-acc} \sim 8 M_\odot$ were accreted and $M_j \sim 4 M_\odot$
blown as a collimated fast wind (CFW, defined as not
well-collimated jets) by the accretion disk.
This CFW mass shaped the $8 M_\odot$ blown by the primary,
mainly toward the polar directions, and not captured by the secondary star.
The total mass in the Homunculus is $M_{\rm GE-h}=12 M_\odot$
{{{ (the calculations in this section can be easily scaled to a higher Homunculus
mass of up to $\sim 20 M_\odot$, as claimed for by Smith \& Ferland 2006). }}}
Taking average radius and mass during the accretion phase of the Great Eruption,
we find the total gravitational energy released by the accreted
mass
\begin{eqnarray}
E_{\rm GE-2}  \simeq \frac{GM_{2}M_{\rm GE-acc}}{R_{2}}
=6 \times 10^{49}
\left( \frac{M_{2}}{30M_\odot} \right)
\left( \frac{M_{\rm GE-acc}}{8M_\odot} \right)
\left( \frac{R_{2}}{15 R_\odot} \right)^{-1}   \erg.
\label{e2}
\end{eqnarray}

The speed of the CFW is about the escape speed from the secondary star
$v_j \sim 1000 \km \s^{-1}$, which amounts to a total energy of
$E_{\rm jet}=4 \times 10^{49} \erg \s^{-1}$ (for a CFW mass
of $M_j=4 M_\odot$), which is more than the present kinetic energy of the
Homunculus, $3 \times 10^{49} \erg$ (for a mass of $M_{\rm GE-h}=12 M_\odot$).
Taking the speed of the primary wind to be $100 \km \s^{-1}$,
the total momentum deposited into the Homunculus is
$(8 \times 100)+(4 \times 1000)=4800 M_\odot \km \s^{-1}$.
This is close to the total momentum of the Homunculus I calculated from the
mass and velocity distribution given by Smith (2006)
$5600 M_\odot \km \s^{-1}$ (for a mass of $12 M_\odot$).
The energy and momentum budget shows that the interaction between the CFW and
the primary mass was between purely momentum conserving and purely kinetic
energy conserving, i.e., some of the initial kinetic energy in the CFW
was radiated away.

The total radiated energy from the Great Eruption is
$E_{\rm GE-rad} = 3 \times 10^{49} \erg$ (Humphreys et al. 1999).
Considering that the primary itself has a luminosity of
$\sim5 \times 10^6 L_\odot$ (Humphreys et al. 1999),
it contributed $E_{\rm GE-1}=1.2 \times 10^{49} \erg$ during
the Great Eruption.
Examining equation (\ref{e2}), one finds that the energy liberated by the
accreted mass can easily account for the extra radiated energy and the kinetic
energy in the Homunculus, which are summed up to $\sim 5 \times 10^{49} \erg$.

In the binary model the extra radiated energy of
$\sim 2 \times 10^{49} \erg$ came from the gravitational energy of the
accreted mass.
Namely, the primary luminosity {\it did not change during the Great Eruption.}.
This suggests that the primary eruption was not much different from S Doradus type
eruptions, where the stellar bolometric luminosity does not
change (Humphreys et al. 1999).
The erupting primary star mainly supplied the mass to the secondary and the
Homunculus, but not the extra radiated energy.

\subsection{Angular Momentum Evolution}
Consider an envelope of a giant star rotating as a solid body.
The dimensionless quantity $\beta$ is defined by taking the angular momentum loss
rate from the envelope to the wind to be
$\dot J_{\rm wind} = \beta \omega R^2 \dot M$,
where $\omega$, is the stellar angular velocity, $R$ the stellar radius,
and $\dot M$ the mass loss rate.
The value of $\beta$ depends on the mass loss geometry:
for a constant mass loss rate per unit area on the surface $\beta =2/3$,
while for an equatorial mass loss $\beta=1$.
The moment of inertia of the envelope is $I = \alpha M_{\rm env} R^2$.

For the mass loss geometry found by Smith (2006) I calculate
that $\beta =0.38$. 
As discussed in Soker (2004) I take $\alpha=0.1$ for the eruptive $\eta$ Car model.
I assume that during the eruption the star continued to rotate as a solid body.
This implies that during the eruption the angular velocity of the star
evolved according to
\begin{eqnarray}
\frac{\omega}{\omega_0}=\left( \frac{M_{\rm env}}{M_{\rm env0}} \right)
^{\frac{\beta}{\alpha}-1} =\left( \frac{M_{\rm env}}{M_{\rm env0}} \right)^{2.8}.
\label{omega}
\end{eqnarray}
As in Soker (2004) I take an envelope mass of about half the stellar mass,
or $M_{\rm env0}=70 M_\odot$ at the beginning of the Great Eruption.
As stated in Soker (2004), this is an upper limit on the envelope
mass; the envelope mass is likely to be lower, increasing the spin
down rate.
The mass lost at the eruption into the Homunculus according to the single
star model is $M_{\rm GE-h} \simeq 12 M_\odot$ (Smith et al. 2003b).
I find that at the end of the eruption ${\omega}/{\omega_0} \simeq 0.6$.
Namely, the star had substantially spun-down.
Even for the extreme case of $\alpha=0.2$ the progenitor of $\eta$ Car
spun down during the Great Eruption.
{{{ This finding, based on the assumption of a solid-body rotation, is contrary }}}
to the claim made by Smith (2006) that the post-outburst star had higher angular
momentum per unit mass than the pre-outburst star.

According to the single star model, it is still possible that the star
was not rotating as a solid body, but rather the angular velocity closer to
the rotation axis was lower than near the equator.
In such a case the specific angular momentum of the star
could have increase for the mass loss geometry found by Smith (2006).
However, there is no model for such a behavior.
For example, if magnetic fields play a role, then, using results of
magnetically active main sequence stars, e.g., the sun, we know that
even with higher mass loss rate along the polar directions these stars
spin down.

Alternatively, we can consider mass loss due to radiation pressure.
During the Great Gruption the stellar envelope was optically thick
to a huge radius (Davidson \& Humphreys 1997). In such a case we would expect
spherical mass loss geometry, as radiation pressure would act to all directions.
Another problem with a polar outflow from the primary star is that
there is not enough radiation energy and momentum to drive the outflow
if only the polar direction contributes to the mass loss (Soker 2004).

Groh et al. (2006) found the LBV AG Carinae to be a fast rotator.
They, as well as Smith (2006), viewed this finding as a support for
models of axisymmetrical wind from rotating stars as the main
shaping process.
However, the behavior of AG Car is opposite to the required behavior during
the Great Eruption.
Groh et al. (2006, Table 2) found that the mass loss rate increases when the star
radius increases, its effective temperature decreases, and its rotation velocity
{\it decreases}.
As the mass loss rate increase by a factor of 2.6 the ratio of rotation to
critical velocity decreases from 0.86 to 0.57.
Namely, the high mass loss rate is expected to become more spherical when
mass loss rate is higher in single star models.
High mass loss rate episode, as the Great Eruption, will not be highly
asymmetrical if the shaping is to result from a behavior as that of AG Car.
In the binary model the launching of jets by the accreting companion
is more likely to occur as the primary wind velocity decreases, as
is expected to be the case when mass loss rate is higher, as in the
Great Eruption (see section 3.2).

In summary,  I {{{ disagree }}} with the claim made by Smith (2006) that a single
star can account for the mass loss geometry he finds, and that the post-outburst
star had higher angular momentum per unit mass than the pre-outburst star.

\section{COMPARING THE RED RECTANGLE WITH $\eta$ CARINAE}
\label{redr}
\subsection {Similar Properties}
Some similarities between $\eta$ Car and PNs and other related objects
were noticed before (Allen \& Swings 1972; Balick 1989; Soker 2001b, 2004 2005b;
Smith 2003; Smith et al. 2005).
{{{ In particular, the detail comparison of the PN M2-9 to $\eta$ Car conducted
by Smith et al. (2005) also shows that such a comparison can shed light
on the evolution of these systems. }}}
The new results of Smith (2006) allows a better comparison between the Homunculus
and the Red Rectangle, the nebula around the binary system HD~44179.
In this section I summarize and extend the list of similarities between the Red Rectangle
and $\eta$ Car.

\subsubsection{Eccentric binary system}
At the center of the Red Rectangle nebula there is the HD~44179 binary system.
This binary system has an orbital period of $T_{\rm orb} = 322$~days,
a semimajor axis of $a \sin i = 0.32 \AU$, and an eccentricity
of $e=0.34$ (Waelkens {\it et al.} 1996; Waters {\it et al.} 1998;
Men'shchikov et al.\ 2002).
Men'shchikov et al.\ (2002) suggest that the orbital separation is
$a=0.9 \AU$, and the masses of the two components are $M_1 \simeq 0.6 M_\odot$
and $M_1 \simeq 0.35 M_\odot$.

Based on several papers (e.g., Ishibashi et al. 1999; Damineli et al. 2000;
Corcoran et al. 2001; Hillier et al. 2001; Pittard \& Corcoran 2002;
Smith et al. 2004), I take the following parameters of $\eta$ Care.
The stellar masses are $M_1=120 M_\odot$, $M_2=30-40 M_\odot$, the eccentricity is
$e=0.9$, and orbital period is $T_{\rm orb} = 2024$~days, hence the semi-major axis is
$a=16.64 \AU$ (for $M_2=30 M_\odot$), and the orbital separation at periastron is
$r=1.66 \AU$.
The mass loss rates are $\dot M_1=3 \times 10^{-4} M_\odot \yr^{-1}$
and $\dot M_2 =10^{-5} M_\odot \yr^{-1}$.
The terminal wind speeds are taken to be $v_1=500 \km \s^{-1}$ and
$v_2=3000 \km \s^{-1}$.

Some similar aspects of the evolution of the two binary systems were discussed
in Soker (2005b).
Here I add that the primary stars in both systems
are evolved stars, and are much larger than their respective sizes on the
main sequence, and both have a relatively high mass loss rate.
Both stars are not far from filing their Roche lobe; both have
$R_1/a_p \simeq 0.3$  where $a_p$ is the separation at periastron.
In the past, episodically the radius of the primary star of both systems
could have been much larger, filling its Roche lobe (see section 3.2).
While the primary in $\eta$ Car is more massive than half its birth mass,
the primary in the Red Rectangle seems to be less than half its birth mass.
Also, while the primary in $\eta$ Car is closed to its Eddington luminosity,
the luminosity of the primary (a post-AGB star) in HD 44179 is $\sim6000 L_\odot$
(Men'shchikov et al.\ 2002), only $\sim 0.3$ of its Eddington limit.

\subsubsection {General Shape}
\begin{figure}
\vskip 2.2 cm
{\includegraphics[scale=0.89]{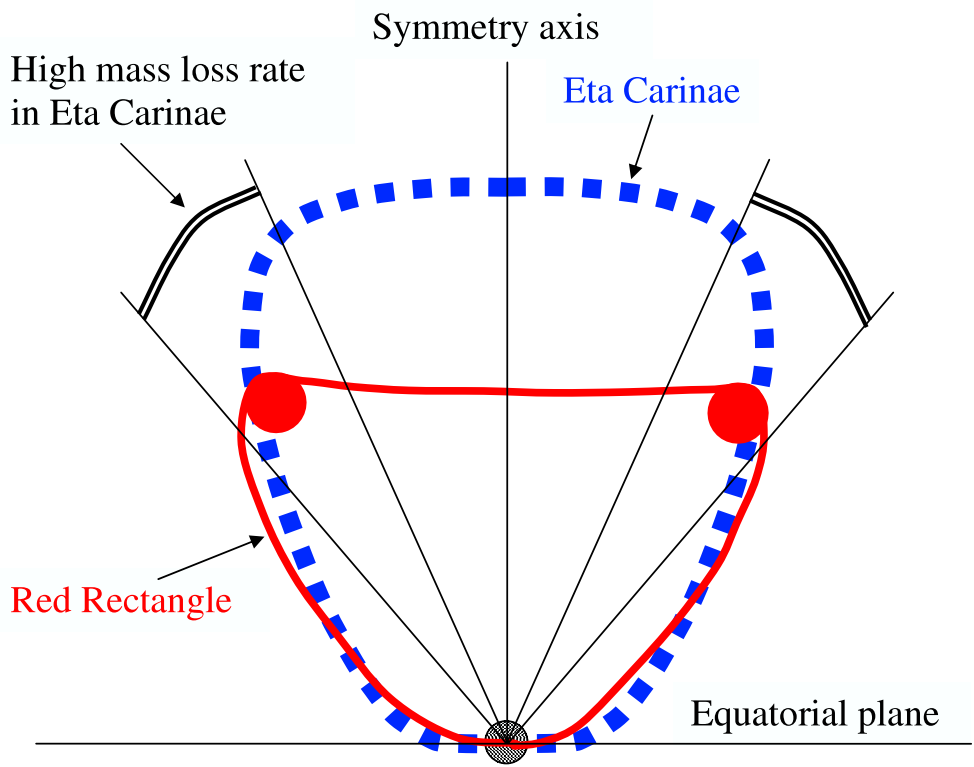}}
\vskip -9.2 cm
\caption{The dashed thick line shows the shape of the dense shell of $\eta$
Car as drawn in Fig. 6 of Smith (2006). The double-arcs angle shows the direction
where mass per solid angle is the maximum: $50^\circ-65^\circ$
(measured from the equatorial plane), according to Fig. 7 of Smith (2006).
The curved solid line with the two symmetric circles show the shape of
one of several `Wineglass' shaped bipolar structure in the Red Rectangle
(from Fig. 10 of Cohen et al. 2004). In the particular example here the
second `Wineglass' from the center in the southern lobe is shown.
Only one half of the symmetry plane through the symmetry axis is shown. }
\label{shape}
\end{figure}

Smith (2006) gives the mass in the Homunculus as function of angle
(measured from the equatorial plane).
There is a high mass concentration in the range $\sim 45-65 ^\circ$,
with a higher concentration in the range $\sim 50-65 ^\circ$, and a
peak at $\sim 53^\circ$.
The maximum in mass per unit solid angle  is in the range $\sim 50-65 ^\circ$.
The shape of the Homunculus as given by Smith (2006) is shown in Figure
\ref{shape} by the dashed-thick line. Also marked are the directions
containing high mass concentration.

The general structure of the inner region of the Red Rectangle
is presented in Figure 10 of Cohen et al. (2004).
Each pair of `vortices'$-$the two large circular spots on Figure \ref{shape}$-$on
opposite side of the symmetry axis are a limb-brightened projection of a single ring.
This ring is connected with a paraboloid to the center.
One such structure, out of the several in the Red Rectangle, is shown in
Figure \ref{shape}.
Such a structure is not unique to the Red Rectangle.
The outer lobe pair of the PN M2-9 has this structure (see HST home page
and Doyle et al.\ 2000).

>From Figure \ref{shape} we can see that both the general structure and the
concentration of mass is similar in the Homunculus and in each mass loss
episode of the Red Rectangle.
The main difference is that in the Red Rectangle there are several bipolar structures,
which are merged to form large biconical structure.
In $\eta$ Car there is only the Homunculus, although the little Homunculus
(Ishibashi et al. 2003) might develop into a similar shape,

\subsubsection{Departure from axisymmetry}

Departure from axisymmetry is observed in $\eta$ Car (e.g., Smith 2002; see
discussion in Soker 2001b), and in the inner region of the Red Rectangle
(e.g., Tuthill et al. 2002; Miyata et al. 2004).
In Soker (2001b) I argued that the departure of the Homunculus
from axisymmetry can be explained by with binary model.

\subsubsection{A slow equatorial flow}

There is a slow, $\sim 100 \km \s^{-1}$, equatorial outflow in $\eta$ Car
(Zethson et al. 1999), called the skirt.
Its total mass is small compared with the Homunculus mass (Smith 2006).
Around HD 44179 there is a cold disk with a size of hundreds astronomical units,
larger than the binary system and smaller than the nebula  (e.g., Jura \& Kahane 1999;
Dominik et al.\ 2003; Bujarrabal et al.\ 2003).
In addition, there is a slow outflow starting from the outer
edge of the disk (Bujarrabal et al.\ 2003).

In Soker (2005b) I compared also the large dust grains observed
in both systems (e.g., Smith et al. 2003b for $\eta$ Car; Cohen et al.
1975 and Tuthill et al. 2002 for the Red Ractangel).
Such large grains can be formed in a dense and slow equatorial outflow
(Soker 2000b; Men'shchikov et al.\ 2002)
or in a long leaved disk (e.g., Jura et al. 1995; Jura \& Kahane 1999).

\subsection {Conclusions from the Similarities}

The comparison made in section 3.1 is meaningful only if
the same basic mechanism is responsible for the
shaping of the Red Rectangle and the Homunculus.
Following previous papers (Soker 2001b, 2004, 2005b) I assume that
this is indeed the case, and come to the following conclusions.

(1) The shaping is not related to the proximity of $\eta$ Car's luminosity
to the Eddington limit.
As stated above, the primary in the Red Rectangle is far from its Eddington limit.

(2) Fast rotation of the primary is not likely to be behind the bipolar
structure. Since the primary in the Red Rectangle lost most of its envelope mass,
it must be rotating very slowly (Soker 2004). Even if it was spun-up
by tidal interaction and reached synchronization with the orbital angular velocity,
the rotation velocity of the primary is $<0.2$ times its break-up rotation velocity.
No rotating-star model predict much deviation from spherical
mass loss in such a case.

(3) In both systems the companion is outside the envelope. This means that the
shaping is done by a companion which avoided a common envelope evolution.
The applicable binary mechanism is one where the companion accretes mass and
launches two jets during high mass loss rate episodes (Soker 2001b, 2005b for $\eta$ Car;
Soker 2005a for the Red Rectangle).

(4) During the high-mass loss rate episodes, mainly in $\eta$ Car,
the primary expanded such that the accretion process is a mixture of accretion
from a wind and a Roche lobe over flow.
I will now show this.

For a pure accretion from a wind, namely, the wind is considered to
be at a large distance from the star and at its terminal speed,
the Bondi-Hoyle accretion process is applicable.
The Bondi-Hoyle accretion radius is
\begin{equation}
R_{\rm acc2}= \frac {2 G M_2}{v_{\rm rel}^{2}}  = 5   
\left( \frac{M_2}{ 30 M_\odot}\right)
\left( \frac {v_{\rm rel}}{100 \km \s^{-1}} \right)^{-2} {\rm AU},
\label{accrad}
\end{equation}
where $v_{\rm rel}$ is the relative velocity between the primary wind and
the secondary.
The present primary's wind terminal speed in $\eta$ Car is $\sim 500 \km \s^{-1}$.
However, I assume it was much slower during the
Great Eruption. To be considered as a free wind, the speed at apastron,
$a_a\simeq 30 \AU$ should exceed the escape velocity from the system
$v_{\rm esc}(30 \AU) \simeq 85 \km \s^{-1}$.
The accretion radius is much smaller than the orbital separation,
and the fraction of the accreted mass from the wind near apastron
would be $\sim (R_{\rm acc2}/2a_a)^2 \sim 0.015$, much less than that
required by the model presented in section 2.
Near periastron the distance is much smaller, but the orbital velocity increases,
and this fraction increases, but only to $\sim 0.25$, and only
for the relatively short time the secondary spends near periastron.

The situation with the Red Rectangle is better (Soker 2005a).
The wind from post-AGB stars is very slow, $\sim 10 \km\s^{-1}$, and
the orbital velocity is larger than the wind's speed.
The relative orbital velocity between the two stars in HD~44179 is $\sim 30 \km \s^{-1}$.
Taking the orbital separation to be $\sim 1 \AU$, I find the fraction of mass
that would be accreted from a free wind to be $\sim 0.5$, as
required by the model (Soker 2005a).

For the companion to blow a jet an accretion disk should be formed.
The condition for the formation of an accretion disk is that
the specific angular momentum of the accreted matter $j_a$ be
larger than the specific angular momentum of a particle in a
Keplerian orbit at the equator of the accreting star of radius $R_2$:
$j_2=(G M_2 R_2)^{1/2}$. When a compact secondary
star moves in a circular orbit at radius $a$ and accretes from the wind of a
mass losing star, such that the accretion flow reaches a steady
state, the condition for the formation of an accretion disk reads
(see Soker 2001a for more detail and previous references)
\begin{equation}
1< \frac {j_a}{j_2} \simeq 0.1
\left( \frac{\eta}{0.2} \right) \left( \frac {M_1+M_2}{150
M_\odot} \right)^{1/2} \left( \frac {M_2}{30 M_\odot}
\right)^{3/2} \left( \frac {R_2}{20 R_\odot} \right)^{-1/2} \left(
\frac {r}{30 \AU} \right)^{-3/2} \left( \frac{v_{\rm rel}}{100
\km \s^{-1}} \right)^{-4}.
\label{jacc}
\end{equation}
$\eta$ is the ratio of the accreted angular momentum to that
entering the Bondi Hoyle accretion cylinder.
In an eccentric orbit near apastron, the value of $j_a$ will be smaller due to
lower azimuthal velocity than in a circular orbit.
I conclude that in the case of pure accretion from the wind, no accretion
disk could have been formed during the Great Eruption of $\eta$ Car.
This suggests that during the Great Eruption the primary swelled,
and its extended envelope overflowed and transferred  mass to the secondary,
increasing substantially both the mass transfer rate and the specific angular
momentum of the accreted mass compared with pure wind accretion.

Substituting typical values for the HD~44179 binary system at the center of the
Red Rectangle I find the coefficient in equation (\ref{jacc}) to be $\sim 10$
for an accreting WD, and $\sim 1$ for an accreting low mass main sequence stars.
Therefore, in the case of the Red Rectangle an accretion disk can be formed
even by pure accretion from a wind.
However, the similarity of the Red Rectangle and $\eta$ Car raises the possibility
that during episodes of high mass transfer rate in the Red Rectangle,
the primary forms an extended envelope, which is overflowing to the companion.

Accretion of more mass at large separations (near apastron)
increases the eccentricity.
The change in eccentricity due to a mass  $\delta M_{\rm tran}$
transferred from the primary to the secondary is given by (Eggleton 2006)
\begin{eqnarray}
\delta e = 2 \delta M_{\rm tran}
\left( {\frac{1}{M_1}} - {\frac{1}{M_2}} \right) (e+\cos \theta),
\label{ecc4}
\end{eqnarray}
where $\theta$ is the orbital angle, with $\cos \theta=1 $ at periastron
and $\cos \theta=-1 $ at apastron
For a mass of $12 M_\odot$ captured by the secondary during
the Great Eruption (out of which $8 M_\odot$
accreted and $ 4 M_\odot$ blown in the jets; see section 2), most
of it near apastron, I find that $\delta e \sim 0.1$ for initial eccentricity of
$e_0=0.8$. Therefore, a few such repeated eruptions (Smith 2007; Smith \& Owocki 2006)
can be responsible for the high eccentricity of $\eta$ Carinae.
The same process can work in the case of HD~44179.

\section{SUMMARY}
\label{summary}

In this paper I continued and updated the comparison of $\eta$ Carinae and
its bipolar nebula$-$the Homunculus$-$with other astrophysical objects
(Soker 2005b), and the comparison of the binary model with single star models
for the shaping of the Homunculus (Soker 2004).
Basically, I strengthened the binary model for the formation of the Homunculus.
In the binary model (Soker 2001b, 2004, 2005b) a large fraction, and even most,
of the mass lost by the primary star during the 20 years Great Eruption was
transferred to the secondary star (the companion).
An accretion disk was formed around the companion, and the companion
launched two opposite jets (or a collimated fast wind).
These jets shape the circumbinary gas to the observed bipolar structure.

The main results are as follows.
\begin{enumerate}
\item The source of the extra energy of the Great Eruption, both the radiated energy
and the kinetic energy in the Homunculus, is easily accounted for by the gravitational
energy of the mass accreted onto the secondary star (section 2.1).
This suggests that the  eruption of the primary star was not much different from
S Doradus type eruptions, where the stellar bolometric luminosity does not
change (Humphreys et al. 1999).
The periodic peaks during the Great Eruption (Damineli 1996) support the
claim of this paper that accretion onto the secondary star supplies the extra energy
in the Great Eruption. I also suggest that other giant eruptions of LBV stars are
caused by mass transfer onto secondary star.
Unlike Humphreys et al. (1999),  I don't find the assumption that the other three LBV
they list as having giant eruptions have binary companions unreasonable.
The claim that the mass loss during the Great Eruption could not have been
redirected toward the poles by deflection from its companion star because of
energy considerations (Smith 2006) is not relevant to the binary model.
\item Using the new mass distribution in the Homunculus reported by
Smith (2006), I find that in the single star models the primary star in $\eta$ Car
must have spun down substantially during the Great Eruption (section 2.2).
This further strengthen earlier claims (Soker 2004) that single star models
encounter severe problems in explaining the bipolar structure
of the Homunculus.
The axisymmetrical structure of the nebulae of many LBV also support the
binary model (Nota et al. 1995).
\item I made a point that the mass distribution in the Homunculus reported by
Smith (2006) is similar to that in the Red Rectangle, a nebula around
the post-AGB binary system HD~44179 (section 3.1).
This is in addition to other similarities between the two systems (Soker 2005b):
central eccentric binary system where the primary is an evolved star;
slow equatorial outflow; and departure of the nebula from axisymmetry.
\item Assuming that the Red Rectangle and the Homunculus share the same
basic shaping process, I reached the following conclusions (section 3.2):
(4.1) The proximity of the luminosity of the primary star in  $\eta$ Car to
the Eddington limit is not related directly to the shaping of the Homunculus.
The luminosity is very likely to be responsible for the large amount
of mass that was lost during the Great Eruption, though.
(4.2) Fast rotation of the primary star in $\eta$ Car is not the cause
of the bipolar structure.
(4.3) The secondary star did not go through a common envelope evolution.
If it did, it was for a short time, and this is not the mechanism behind the
shaping of the Homunculus. The shaping took place while the secondary was mainly
outside the envelope.
(4.4) The mass transfer mode was more like a Roche lobe overflow than accretion
from a wind. At the beginning of the Great Eruption the primary formed
an extended envelope, to a radius of $\sim 20 \AU$. This envelope
overflowed and transferred mass to the secondary.
Most of the mass transfer occurred at large orbital separations (near apastron),
increasing the eccentricity of the binary system.
\end{enumerate}

{{{{ The results listed above are not extremely sensitive to the parameters
of neither $\eta$ Car nor of the Red Rectangle.
The conclusions are only based on the presence of a companion close enough to accrete
mass and angular momentum at a hight rate from the primary wind.
The companion must be outside the primary envelope in order to blow the jets.
Over all, the periastron distance should be $\sim 1.5-3$ times larger than
the primary-stellar radius when mass transfer occurs.
What is sensitive to the parameters of $\eta$ Car and the Red Rectangle is the exact
shape of the bipolar nebula.
For example, the ratio of the radiative cooling time in the jet and the flow time of
jet determines the shape of the bubble (lobe) inflated by the jet.
Fast and low density jets have a long radiative cooling time,
and are more likely to form `fat' lobes.
The many shapes of PNs show that many types of nebulae can be formed by binary systems.
If $\eta$ Car will go again through a great eruption, it will again form
a bipolar structure composed of two lobes, but the detail of the structure be different.
The departure from axisymmetry is somewhat sensitive to the jet launching period and to the
eccentricity. For example, long active jets launched in a
circular-orbit binary system will not lead to a departure from axisymmetry.
}}}}

What could have causes the primary to lose a huge amount of mass, $\sim 20 M_\odot$,
in only $\sim 20 \yr$ of the Great Eruption?
It seems that nuclear eruptions in the core, as well as other eruptive events
in the core, can be ruled out from what we know about AGB stars.
Helium shell flashes (thermal pulses) on the outskirts of cores of AGB
stars reach luminosities of $>10$ times the normal nuclear luminosity
in the core, but the surface luminosity does not change, or even decreases
(e.g., Iben \& Renzini 1983). Most of the liberated energy goes to
lift mass in layers deep in the envelope.
The disturbance should take place close to the stellar surface.
I speculate that magnetic activity is behind the Great Eruption of $\eta$ Car.

Soker (2000a) and Garc\'{\i}a-Segura et al. ( 2001) propose magnetic activity
cycles to explain the semi-periodic
enhanced mass loss rate episodes observed in progenitors of some PNs
and Proto-PNs (e.g., Hrivnak et al. 2001) and around AGB stats
(e.g., Mauron \& Huggins 2000).
The time intervals between consecutive ejection events are ~200-1000 yr
(Hrivnak et al. 2001).
I take the high mass loss rate into each shell to last several$\times 10 \yr$,
and the mass loss rate to be $\sim 10 \dot M_n$, where
$\dot M_n \simeq 10^{-5} M_\odot \yr^{-1}$ is the normal mass loss rate at
the end of the AGB.
I crudely find that a plausible value for the mass in each shell might be
$M_s \sim 3 \times 10^{-3} M_\odot$. With typically 20 shells at the last
$\sim 10^4 \yr$ of the AGB, I find that $\sim 0.05 M_\odot$ is lost
in the shells.

Interestingly, the estimated mass in each of the semi-periodic concentric shells
(arcs) around AGB stars, $M_s \sim 3 \times 10^{-3} M_\odot$, is of the order of
the mass residing above the convective region of AGB stars, namely, in the
outer radiative zone.
Using the evolving AGB stellar model of Soker \& Harpaz
(1999)\footnote{Note that the density scale in Figs.\ 1-5 of
Soker \& Harpaz (1999) is too low by a factor of 10;
the correct scale is displayed in their Fig. 6.},
I find that when the envelope mass is $M_{\rm env}= 0.3,  0.1,  0.03 M_\odot$
(the effective temperature at these three evolutionary points is $2700,  2700, 3100 \K$)
the mass in the outer radiative zone is $M_{\rm rad} = 0.007, 0.005, 0.003 M_\odot$.
This suggests that during eruptive phases AGB stars expel a mass
about equal to that in the outer radiative zone.
A magnetic activity on the outer boundary of the convective region might
disturbed this region, and leads to its expulsion.

The mass estimated to be ejected by $\eta$ Car and similar
very luminous stars is $\sim 10-15 M_\odot$ (Smith \& Owocki 2006).
In the binary model some mass is accreted by the secondary, so this
mass can be as large as $\sim 20 M_\odot$.
This is of the order of the mass residing in the outer radiative zone
of very massive ($M_1 \ga 80 M_\odot$) stars when they are at the same stage
as $\eta$ Car is (Meynet \& Maeder 2003, 2005).
Namely, the age is few$~10^6 \yr$, effective temperature is $\sim 20,000-25,000 \K$,
and luminosity $\ga 10^6 L_\odot$.
I speculate that the Great Eruption was an event where magnetic activity in the
convective region strongly disturbed the outer radiative zone, and formed an
extended envelope.
One way to formed an extended envelope is by the contraction
of the inner layers.
Thus, it is possible that the magnetic activity in the convective zone caused,
in an yet unprescribed process, the contraction of the convective envelope.
As a result the radiative zone above it expanded to huge dimensions.
Radiation pressure aided in uplifting the mass in the radiative zone and
formed the extended envelope, up to $\sim 20 \AU$, and in expelling mass.

I thank Nathan Smith for useful comments.
This research was supported by grant from the
Asher Space Research Institute at the Technion.


\begin{references}
\reference{}  Allen, D. A., \& Swings, J. P. 1972, ApJ, 174, 583

\reference{} Balick, B. 1989, AJ, 97, 476

\reference{} Balick, B. 2000, in ASP Conf. Ser., Asymmetrical Planetary Nebulae II :
>From Origins to Microstructures, ed. J. Kastner, S. Rappaport, \& N.
Soker (San Francisco: ASP), 41

\reference{} Blanton, E. L., Sarazin, C. L., McNamara, B. R., \&
     Wise, M. W. 2001, ApJ, 558, L15

\reference{}  Bujarrabal, V., Neri, R., Alcolea, J. \& Kahane, C.
     2003, A\&A, 409, 573

\reference{} Cohen, M. et al. 1975, ApJ, 196, 179 

\reference{} Cohen, M., Van Winckel, H., Bond, H. E., \& Gull, T. R.
      2004, AJ, 127, 2362

\reference{} Corcoran, M. F., Ishibashi, K., Swank, J. H., \&
  Petre, R., 2001, ApJ, 547, 1034

\reference{} Corradi, R. L. M., Livio, M., Balick, B., Munari, U.,
    \& Schwarz, H. E. 2001, ApJ, 553, 211

\reference{} Corradi, R. L.  M. , \& Schwarz, H.  E.
    1995, A\&A, 293, 871

\reference{} Damineli, A., 1996, ApJ, 460, L49

\reference{} Damineli, A., Kaufer, A., Wolf, B., Stahl, O.,
    Lopes, D. F., \& de Araujo, F. X. 2000, ApJ, 528, L101

\reference{} Davidson, K., \& Humphreys, R. M. 1997, ARA\&A, 35, 1

\reference{} {{{ Davidson, K., Smith, N., Gull, T. R., Ishibashi, K.,
     \& Hillier, D. J. 2001, AJ, 121, 1569 }}}

\reference{} Dominik, C., Dullemond, C. P., Cami, J., \&
  van Winckel, H. 2003, A\&A, 397, 595

\reference{} Doyle, S., Balick, B., Corradi, R. L. M., \& Schwarz, H. E.
     2000, AJ, 119, 1339

\reference{} Dwarkadas, V. V., \& Balick B., 1998, AJ, 116, 829

\reference{} Dwarkadas, V. V., \& Owocki, S. P. 2002, ApJ, 581, 1337

\reference{} Eggleton, P. 2006, Evolutionary Processes in Binary and Multiple Stars,
   Cambridge University Press (Cambridge)


\reference{} Fabian, A. C., Celotti, A., Blundell, K. M.,
    Kassim, N. E., \& Perley, R. A. 2002, MNRAS, 331, 369

\reference{} Garc\'{\i}a-Segura, G., Lopez, J. A., \& Franco, J.  2001, ApJ, 560, 928

\reference{} Gonzalez, R. F., de Gouveia Dal Pino, E. M., Raga, A.
C., Velazquez, P. F. 2004, ApJ, 600, L59

\reference{} Groh, J. H., Hillier, D. J., \& Damineli, A. 2006, ApJ, 638, L33

\reference{}  Guerrero, M. A., Chu, Y.-H., Manchado, A., \&
    Kwitter, K. B. 2003, AJ, 125, 3213

\reference{} Hillier, D. J., Davidson, K., Ishibashi, K., \& Gull, T. 2001, ApJ, 553, 837

\reference{} Hrivnak, B. J., Kwok, S., \& Su, K. Y. L. 2001, AJ 121, 2775

\reference{} Humphreys, R. M., Davidson, K., \& Smith, N. 1999, PASP, 111, 1124

\reference{} Iben, I., Jr. \& Renzini, A.  1983, ARA\&A, 21, 271

\reference{} Ishibashi, K. et al. 2003, AJ, 125, 3222  

\reference{} Ishibashi, K., Corcoran, M. F., Davidson, K., Swank, J. H.,
  Petre, R., Drake, S. A., Damineli, A., \& White, S. 1999, ApJ, 524, 983

\reference{} Jura M., Balm, S. P., \& Kahane, 1995, ApJ, 453, 721

\reference{} Jura M., \& Kahane C., 1999, ApJ, 521, 302

\reference{} Langer, N., Garc\'{\i}a-Segura, G., \& Mac Low, M.-M
  1999, ApJ, 520, L49

\reference{} Maeder, A., \& Desjacques, V. 2001, A\&A, 372, L9

\reference{} Matt, S., \& Balick, B. 2004, ApJ, 615, 921

\reference{} Mauron, N., \& Huggins, P. J. 2000, A\&A, 359, 707

\reference{} McNamara, B. R., et al.\ 2000, ApJ, 534, L135

\reference{} Men'shchikov, A. B., Schertl, D., Tuthill, P. G.,
   Weigelt, G., \& Yungelson, L. R. 2002, A\&A, 393, 867

\reference{} Meynet, G. \& Maeder, A. 2003, A\&A, 404, 975
\\(for detail see http://obswww.unige.ch/~dessauges/evol/results.html)

\reference{} Meynet, G. \& Maeder, A. 2005, A\&A, 429, 581
\\(for detail see http://obswww.unige.ch/~dessauges/evol/results.html)

\reference{} Miyata, T., Kataza, H., Okamoto, Y. K., Onaka, T., Sako, S.,
  Honda, M., Yamashita, T., \& Murakawa, K. 2004, A\&A 415, 179

\reference{} Morris, M. 1987, PASP, 99, 1115

\reference{} Morse, J. A., Davidson, K., Bally, J., Ebbets, D., Balick, B., \&
    Frank, A. 1998, AJ, 116, 2443

\reference{} Nota, A., Livio, M., Clampin, M., \& Schulte-Ladbeck, R. 1995, ApJ, 448, 788

\reference{} O'Hara, T. B., Meixner, M., Speck, A. K., Ueta, T., \& Bobrowsky, M. 2003, ApJ, 598, 1255

\reference{} Pittard, J. M., \& Corcoran, M. F 2002, A\&A, 383, 636

\reference{} Smith, N. 2002, MNRAS, 337, 1252

\reference{} Smith, N. 2003, MNRAS, 342, 383

\reference{} Smith, N. 2006, ApJ, 644, 1151

\reference{} Smith, N. 2007, in  Massive Stars: From Pop III and GRBs to the Milky Way
     ed. M. Livio  (proceedings of STScI May Symposium 2006) (astro-ph/0607457)

\reference{} {{{ Smith, N., Balick, B., \& Gehrz, R. D.  2005, AJ, 130, 853 }}}

\reference{} Smith, N., Davidson, K., Gull, T.R., Ishibashi, K.,
   \& Hillier, D.J. 2003a, ApJ, 586, 432 

\reference{} {{{ Smith, N., \& Ferland, G. J. 2006, ApJ, in press
(astro-ph/0610530) }}}

\reference{} Smith, N., Gehrz, R., D., Hinz, P. M., Hoffmann, W. F., Hora, J. L., Mamajek, E. E.,
    \& Meyer, M. R. 2003b, AJ, 125, 1458

\reference{} Smith, N., Morse, J. A., Collins, N. R., \& Gull, T. R.
    2004, ApJ, 610, L105

\reference{} Smith, N., \& Owocki, S. P. 2006, ApJ, 645, L45

\reference{} Soker, N. 2000a, ApJ, 540, 436 

\reference{} Soker, N. 2000b, MNRAS, 312, 217 

\reference{} Soker, N. 2001a, MNRAS, 324, 699 

\reference{} Soker, N. 2001b, MNRAS, 325, 584  

\reference{} Soker, N. 2004, ApJ, 612, 1060  

\reference{} Soker, N. 2005a, AJ, 129, 947 

\reference{} Soker, N. 2005b, ApJ, 619, 1064   

\reference{} Soker, N., \& Bisker, G. 2006, MNRAS, 369, 1115

\reference{} Soker, N., \& Harpaz, A. 1999, MNRAS, 310, 1158

\reference{} Soker, N., \& Livio M. 1994, ApJ, 421, 219

\reference{} Soker, N., \& Rappaport, S. 2000, ApJ, 538, 241

\reference{} Tuthill, P. G., Men'shchikov, A. B., Schertl, D., Monnier, J. D., Danchi, W. C.,
  \& Weigelt, G.  2002, A\&A, 389, 889


\reference{} van Boekel, R. et al.\ 2003, A\&A, 410, 37

\reference{} Waelkens, C., Van Winckel, H., Waters, L. B. F. M.,
\& Bakker, E. J. 1996, A\&A, 314, L17.

\reference{} Waters, L. B. F. M. {\it et al.} 1998, Nature, 391, 868

\reference{}  Zethson, T., Johansson, S., Davidson, K.,
  Humphreys, R. M.. Ishibashi, K., \& Ebbets, D. 1999, A\&A, 344, 211


\end{references}
\end{document}